\documentclass[12pt,preprint]{aastex}

\shortauthors{Stanford et al.}
\newcommand{\wcen}{$\omega$ Cen}
\newcommand{\feh}{[Fe/H]}

\newcommand{\eq}{$\approx$}

\begin{document}

\title{A Sr-Rich Star on the Main Sequence of Omega Centauri}

\author{Laura M. Stanford, G. S. Da
 Costa and John E. Norris}
\affil{Research School of Astronomy and Astrophysics, Australian
  National University, Weston, ACT, 2611, Australia}
\email{stanford, gdc, jen@mso.anu.edu.au}
\and

\author{Russell D. Cannon} \affil{Anglo-Australian
Observatory, P.O. Box 296, Epping, NSW, 2121, Australia}
\email{rdc@aao.gov.au}

\begin{abstract}
Abundance ratios relative to iron for carbon, nitrogen, strontium and
barium are presented for a metal-rich main sequence star
\mbox{([Fe/H]=--0.74}) in the globular cluster $\omega$ Centauri.
This star, designated 2015448, shows depleted carbon and solar
nitrogen, but more interestingly, shows an enhanced abundance ratio of
strontium \mbox{[Sr/Fe]\eq1.6~dex}, while the barium abundance ratio
is \mbox{[Ba/Fe]$\leq$0.6~dex}.  At this metallicity one usually sees
strontium and barium abundance ratios that are roughly equal.
Possible formation scenarios of this peculiar object are considered.
\end{abstract}


\section{Introduction}

The largest globular cluster associated with our Galaxy, $\omega$
Centauri, shows substantial ranges in abundance of all elements that
are studied (e.g. \citealt{nd95a, scl95, smi00, pan02}).  The iron
abundance ranges from \mbox{\feh$\approx$--2.0} up to
\mbox{\feh\eq--0.4}.  A peak at \feh=--1.7 exists which accounts for
approximately 70\% of the population, and there is a long tail to
higher metallicities \citep{nfm96, sk96, lee99, pan00, sta06a}.

Studies of individual elements have also shown a range in their
abundances within the cluster members.  Carbon, nitrogen and oxygen
show large scatters in their abundance ratios for a given
\feh~\citep{per80, bw93, nd95a}.  The $\alpha$ elements (Mg, Si, Ca
and Ti) show a constant value of \mbox{[$\alpha$/Fe]{\eq}0.3} below
\mbox{[Fe/H]$<$--1.0} \citep{bw93, scl95, nd95a, smi00} but decrease
to \mbox{[$\alpha$/Fe]=0.0} at higher metallicities
\mbox{(\feh$>$--1.0)} \citep{pan02}. The constancy with iron at lower
abundances indicates they are produced by the same source, most likely
supernovae Type II.  At the higher metallicities, the [$\alpha$/Fe]
decreases, consistent with supernovae Type Ia contributions.

Sodium and aluminium abundance ratios are correlated, and both are
anticorrelated with [O/Fe] \citep{bw93, nd95a, nd95b, smi00}.
\citet{smi00} also report [Al/Fe] being anticorrelated with [Mg/Fe].
[Cu/Fe] has been found to be constant for \mbox{[Fe/H]$<$--0.8}
\citep{smi00, cun02}, but \citet{pan02} reported a trend of increasing
[Cu/Fe] as the metallicity increased from \mbox{\feh=--1.2} to~--0.5.
Studies of the iron peak elements (Cr, Ni, Ti) and metals (Sc, V) have
shown that there is no trend with respect to iron (up to
\mbox{\feh\eq--0.8)}, consistent with primordial enrichment from Type
II supernovae.  The heavy neutron capture elements have been shown by
\citet{nd95a} and \citet{scl95} to increase sharply as a function of
iron abundance.  These results are in contrast to normal globular
clusters, and suggest that the stellar winds from AGB stars (sources
of s-process elements) were involved with the enrichment of {\wcen}.

The cluster also has several members that are of distinct classes,
such as CH stars, Ba stars, S stars \citep{le83} and stars with strong
CO \citep{per80} to name a few.  Here we present another object that
has unusual abundance ratios, the formation of which is difficult to
explain.

We have obtained observations of a sample of main sequence and turnoff
(MSTO) stars within the cluster for several purposes.  Firstly, the
possibility of an age-metallicity relation within the cluster's member
stars was investigated \citep{sta06a}.  We found an age range of
\mbox{2--4} Gyrs exists between the most metal-poor and metal-rich
populations within the cluster, similar to that found by other studies
\citep{hil04, rey04, sol05}. Positions, photometry, metallicities and
ages of the stars in the catalogue can be found in the electronic
version of \citet{sta06a}.  The second goal was to determine abundance
ratios of several elements --- carbon, nitrogen, strontium and barium
as functions of [Fe/H] \citep{sta06b}. In that analysis a metal-rich
main sequence (MS) member, 2015448, was found that exhibited unusual
s-process abundance ratios. The purpose of this Letter is to detail
the abundance analysis for star 2015448, and to propose possible
formation scenarios.  \S \ref{OR_sect} briefly describes the
observation and reduction process for our sample of main sequence and
turnoff members. The stellar parameters and analysis are detailed in
\S \ref{SPA_sect}, while \S \ref{D_sect} discusses the possible
formation scenarios for this object.

\section{Observations} \label{OR_sect}

\begin{figure}
\includegraphics[scale=.60,angle=0]{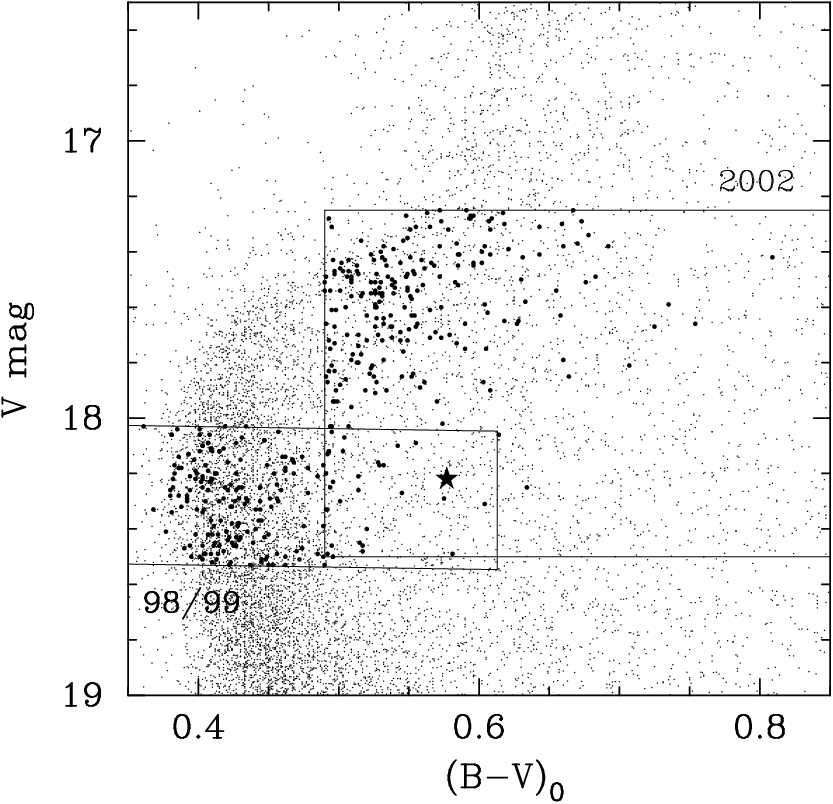} 
\caption{CMD of
$\omega$ Cen sample showing the photometry for the cluster (small
dots), photometry for radial velocity members (large dots) and the
object with unusual abundance ratios --- 2015448 (large star).  The
boxes represent the two regions that were observed in 1998 and 1999,
and 2002. \label{cmd}}
\end{figure}

\begin{figure}
\includegraphics[scale=.85,angle=0]{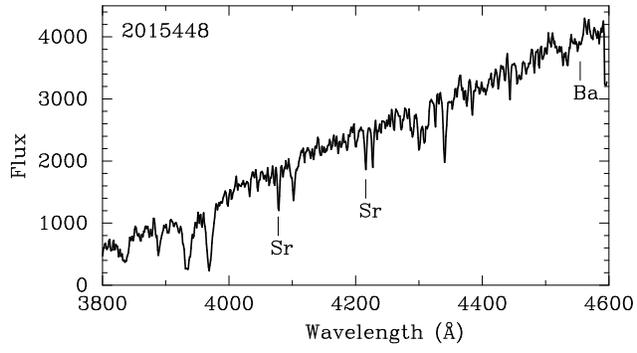} 
\caption{Combined observed spectra of the star 2015448.
\label{spec}}
\end{figure}

The observations and reduction of our data are described in detail in
\citet{sta06a} to which we refer the reader.  Briefly, photometry for
the cluster was obtained in the V and B bands, and samples were chosen
within an annulus \mbox{15'---25'} from the cluster center.  Two
regions were defined near the MSTO, shown in Figure \ref{cmd}, and
spectra were obtained for objects within these CMD regions using the
Two Degree Field Spectrograph (2dF) on the Anglo-Australian Telescope
\citep{lew02}.  Figure \ref{cmd} shows the color-magnitude diagram of
all objects with no membership information as small dots, the sample
of 420 radial velocity members as large dots, and the object that is
the topic of this Letter as a large star.  This object, 2015448, has
$V=18.22$, and $B-V=0.69$.  This star immediately stood out in a
visual inspection of the spectra, due to the anomalously strong Sr
lines, seen in a Figure \ref{spec}.  Its position is
\mbox{RA=13$^h$24$^m$49.10$^s$} and
\mbox{Dec=--47$\degr$40$'$34.7$''$} (J2000).  With a radial velocity
of \mbox{228$\pm$11~kms$^{-1}$}, star 2015448 is likely to be a member of
the cluster as {\wcen} has a radial velocity of
\mbox{232$\pm$0.7~kms$^{-1}$} \citep{din99}

The observations were carried out in half hour exposures, and, as
several such exposures were needed to obtain the required
signal-to-noise of $\sim$30 for each star, there are several
individual observations for star 2015448.  Also, this object was observed
in observing sessions in 1998, 1999 and 2002, in different fibres and
spectrographs. This gives an independent check on the relative
strengths of the Sr and Ba features.  It was found that the two Sr
features were present in all observations, while Ba was not clearly
detected in any of our spectra.

A spectroscopic abundance analysis of the full sample is detailed in
\citet{sta06b}.  Abundances relative to iron of C, N and Sr were
determined using spectrum synthesis techniques.  The CH feature at
$\sim$4300{\AA} was used to determine the [C/Fe] abundance for each
star.  This [C/Fe] was then adopted when the CN feature at 3883{\AA}
was analyzed to find [N/Fe].  The Sr abundance ratio was determined
using the two Sr lines at 4077.71{\AA} and 4215.52{\AA}.  When
possible, [Ba/Fe] was also determined from the feature at
4554.03{\AA}.

\section{Stellar Parameters and Analysis} \label{SPA_sect}

The process for calculating metallicities is described in detail in
\citet{sta06a}.  The Sr-rich star has a metallicity of
\mbox{[Fe/H]=--0.74$\pm$0.23~dex}, determined solely from the
Auto-Correlation Function method, and an age of
\mbox{14.5$\pm$2.2~Gyrs}.  This age is consistent with the mean age of
the bulk of the stars in the cluster.  The ages for the individual
stars in the sample were calculated by fitting an isochrone to each
object using the magnitude, color, metallicity and [$\alpha$/Fe]
abundance (Y$^2$, \citet{yi01}, Green Table).  This also enabled an
individual temperature and gravity to be assigned to each star.  A
temperature of 5820 (K), and \mbox{log$g$=4.2~(cgs)} were adopted for
this object.  A reddening \mbox{$E(B-V)$=0.11} \citep{lub02} and
distance modulus \mbox{(m--M)$_{V}$=14.10} were assumed.  Although the
star's color is redder than the bulk of the population in the cluster,
it is also more metal-rich than that population. The hydrogen line
strengths in the observed spectrum are consistent with those expected
on the basis of the star's color.

Synthetic spectra were obtained using the stellar models of Kurucz
(1993), atomic line lists of Bell (2000, private communication) and
molecular line lists of Kurucz. The spectrum synthesis code was
developed by \citet{cn78}.  More details of this process are given in
\citet{sta06b}.

\subsection{Abundance Ratios} \label{A_sect}

\begin{figure}
\includegraphics[scale=.70,angle=0]{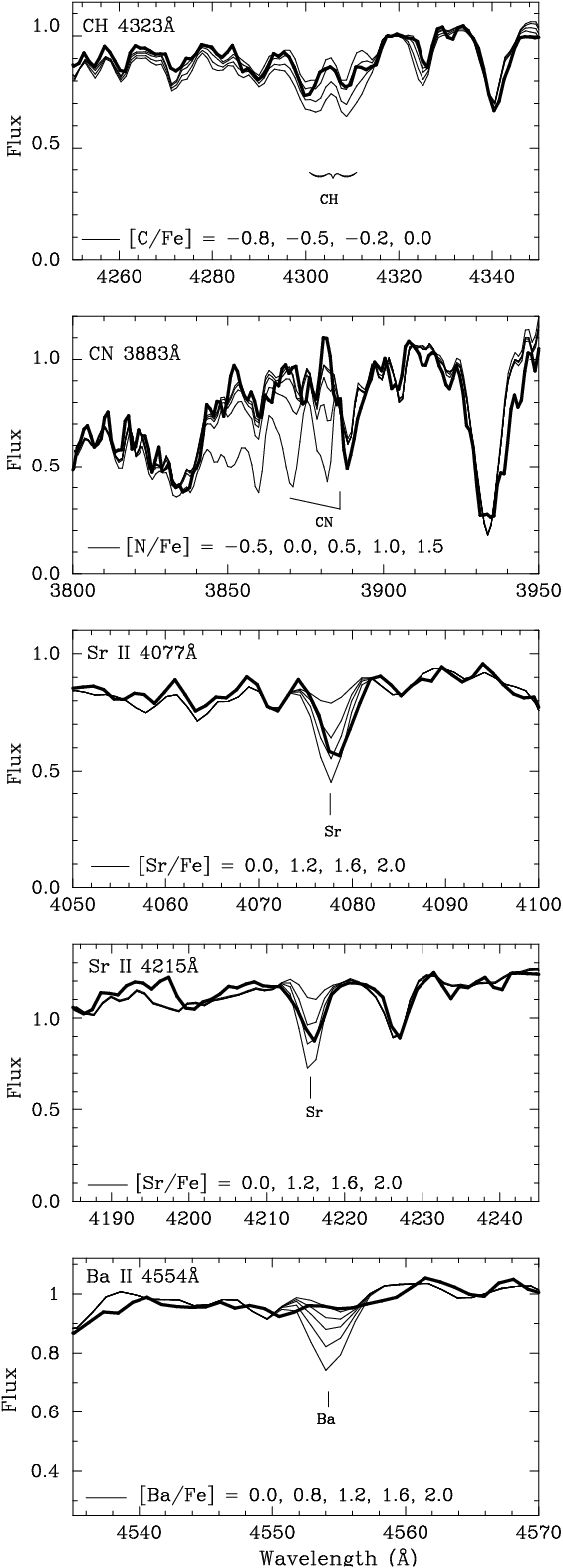} 
\caption{Observed
spectrum (dark line) of the star 2015448 and synthetic spectra (light
lines) in the region of the CH $\sim$4300{\AA}, CN $\sim$3883{\AA}
bands, and Sr {\sc II} 4077{\AA}, Sr {\sc II} 4215{\AA} and Ba {\sc
II} 4554 {\AA} lines.
\label{spec1}}
\end{figure}

Figure \ref{spec1} shows the synthetic spectrum fits to the CH, CN, Sr
and Ba features.  The G band at $\sim$4300{\AA} was analyzed first,
and gave a C abundance of \mbox{[C/Fe]=--0.5$\pm$0.3~dex}.  This
feature loses its sensitivity for abundances less than
\mbox{[C/Fe]=--0.5}, shown by the close proximity of the
\mbox{[C/Fe]=--0.8} and --0.5~dex synthetic lines.  As we have no
information on O, which affects the C abundance obtained from the CH
feature, a value of \mbox{[O/Fe]=0.18~dex} was assumed.  This star was
assumed to have a \mbox{[C/Fe]=--0.5}, which was then used when
determining [N/Fe]. Synthesis of the CN feature at $\sim$3883{\AA} led
to a N abundance ratio of \mbox{[N/Fe]$<$0.5~dex}.  A range of N
abundance ratios are shown in the synthetic spectra, enabling an upper
limit on to be placed on the N abundance ratio.  The CN feature in
Figure \ref{spec1} shows that abundance ratios of \mbox{[N/Fe]=1.5}
and 1.0 are too high.  However, this feature loses its sensitivity as
smaller N abundance ratios are considered and only an upper limit can
be determined.

The Sr and Ba lines were analyzed together.  There were two Sr lines
used, Sr {\sc} II at 4077.71{\AA} and Sr {\sc II} at 4215.52{\AA}.
The Ba {\sc II} 4554.03{\AA} line was used to constrain the Ba
abundance ratio.  The two Sr lines are in agreement and yield
\mbox{[Sr/Fe]=1.6$\pm$0.1~dex}.  It can be seen clearly that a solar
Sr abundance ratio does not fit the observed spectrum.  The CN
bandhead at $\sim$4216{\AA} can affect the abundance ratio obtained
for the Sr 4215{\AA} feature. However with an assumed solar N
abundance ratio, there is little effect to the Sr line.  While the N
abundance ratio could be as high as \mbox{[N/Fe]=0.5~dex}, spectrum
synthesis calculations show this does not affect on the Sr 4215{\AA}
line significantly, and does not alter the abundance ratio obtained.
Using the same enhancement for Ba that was found for Sr, one finds the
predicted strength to be too high to fit the observed Ba line.  The
sensitivity of the Ba feature is low up until abundance ratios of
\mbox{[Ba/Fe]=0.6~dex}, and an upper limit can be placed at this
value.

\subsection{Errors} \label{E_sect}

The stellar parameters temperature, gravity and metallicity, for
star 2015448 were varied individually by their uncertainties to give an
estimate of the error in our determined abundance ratios.  These were
then added in quadrature.  The error in temperature comes from
$\Delta(B-V)$ and the reddening, equating to $\pm$100K uncertainty in
temperature.  This propagated to errors of $\pm$0.2~dex in C and
$\pm$0.1~dex in Sr.  Gravity was changed by $\pm$0.2~dex, and did not
lead to any significant errors in the final abundance ratios.  The
uncertainty in metallicity propagated to errors in abundance ratios
for \mbox{$\Delta$C$\pm$0.2~dex} and \mbox{$\Delta$Sr$\pm$0.1~dex}.
The CN feature used to determine N, adopting the previously determined
C abundance ratio for 2015448, did not show any change with the above
changes in temperature, gravity, metallicity, or the determined error
in the C abundance ratio, due to the low sensitivity of this feature.
To determine the effect of the assumed oxygen abundance,
\mbox{[O/Fe]=0.0} and +0.3 were used and the CH feature reanalyzed.
It was found to have no noticeable effect on the [C/Fe] abundance
determined.

\section{Discussion} \label{D_sect}

This peculiar star in {\wcen} shows \mbox{[C/Fe]=--0.5},
\mbox{[N/Fe]$<$0.5}, enhanced \mbox{[Sr/Fe]=1.6}, and
\mbox{[Ba/Fe]$<$0.6}.  The formation process that created it was
unusual as there are no similar objects yet found within the cluster.
We do find MS objects with high Sr, but these also have similar
enhancements in Ba.  This star was one out of 420 stars studied on the
main sequence turnoff.  Given that it is a metal-rich star, and we
only found 25 such objects, a more extensive search of other areas of
the center of the cluster may prove fruitful in determining if there
are more of these objects.  On the RGB, at a metallicity of
\mbox{\feh=--0.8}, most stars have \mbox{[Ba/Fe]\eq+0.6}
\citep{nd95a}, which is consistent with the upper limit found for
star 2015448.  \citet{nd95a} did not observe Sr, and a direct comparison
cannot be made.  However, they did investigate other light s-process
elements such as Y and Zr.  Both of these elements have abundance
ratios equal to or less than 0.6, and no stars on the RGB exhibit
s-process abundances as high as is found here for [Sr/Fe].

There are at least two possibilities that lead to the formation of
this star. Firstly, it may have been in a binary system with a
companion that underwent unusual s-process enrichment and transfered
mass to the star we see today.  At present we have no evidence that
this star is in a binary system: the individual velocities from the
1998, 1999 and 2002 observations are consistent to within the
measurement errors of \mbox{$\sim$10kms$^{-1}$}.  A second possibility
is that the Sr-rich star formed out of already enriched material.

AGB stars of low and intermediate mass dredge up $^{12}$C to the
surface via helium shell thermal pulses.  In general, if the star is
more massive than $\sim$3M$_{\odot}$, the $^{12}$C is processed to
$^{14}$N via the CN cycle \citep{vdm02}.  The main s-process is
thought to occur within the $^{13}$C pocket where the neutron source
is the \mbox{$^{13}$C($\alpha$,n)$^{16}$O} reaction
(e.g. \citealt{lk01}).  To generate the abundance ratio seen here it
would seem necessary that the s-processing occurs only for the light
s-process elements, perhaps up to the peak at Zr, rather than
progressing through to heavy s-process elements which include Ba.
This was may be due to the neutron density not being high enough, as
higher neutron densities, assuming all other parameters being equal,
will produce larger amounts of heavier nuclei such as Ba or La,
relative to the lighter ones \citep{smi05}.

Another source of s-processing, known as the weak s-process, occurs in
massive stars \citep{phn90}. The \mbox{$^{22}$Ne($\alpha$,n)$^{25}$Mg}
reaction is the source of the neutrons.  Most of the neutrons produced
from this reaction are captured by the light elements, and only a
small fraction are captured by the $^{56}$Fe seed nucleus, a process
known as ``self-poisoning''. This is the reason for the limited
efficiency of the \mbox{$^{22}$Ne($\alpha$,n)$^{25}$Mg} source for
s-process. It allows for the production of only light s-process nuclei
with mass numbers \mbox{65$<$A$<$90} and Sr, with a mass number of 87,
falls into this range. Ba, on the other hand with mass number 137,
does not and little of this element is produced by this process
\citep{phn90}. It is unclear whether massive stars could also produce
the necessary C and N abundance pattern found here, or whether star
2015448 contains other elemental abundance patterns that may have
originated from these objects.

The r-process occurs in an environment that is rich in neutrons, and
the mean time between successive neutron captures in very short
compared with the time to undergo $\beta$-decay.  This scenario as the
enrichment source also requires the presence of SNe products such as
calcium and iron, or to have the Sr enriched material transferred but
not any SNe products.

Using the abundance yields for the r and (main) s-process (occurring
in low-to-intermediate mass AGB stars) from \citet{cam82}, and
normalizing to [Sr/Fe] one finds a difference greater than one dex
between predicted and observed abundances for Ba.  For the main
s-process, the predicted abundance is \mbox{[Ba/Fe]=1.68}, while the
r-process predicts \mbox{[Ba/Fe]=1.84}.  Both of these values are
obviously too large, confirming the unusual nature of the source of
the abundance patterns found in star 2015448.  However, it should be
noted that AGB star yields do not explain the abundance anomalies of
Na, Mg, Al, and O found in other stars of {\wcen} or globular clusters
(see e.g. \citealt{fen04}).  This may be due to the inadequacy of the
models used to calculate yields for these objects.

From weak s-process yields in massive stars \citep{phn90}, using an
initial metallicity of \mbox{Z/Z$_{\odot}$=10$^{-1}$} and mass of
16M$_{\odot}$, \mbox{[Sr/Fe]\eq1.8} and \mbox{[Ba/Fe]\eq0.7}, in good
agreement with the Sr abundance and upper limit of the Ba one obtained
here.  This indicates the weak s-process is the more favorable option
for the source of enrichment in star 2015448.

This star is also highly unusual when compared to observations of
field stars.  Studies of n-capture elements in samples of field stars
for a range of metallicities show a spread in abundance ratios of Sr
and Ba at low metallicities \mbox{([Fe/H]$<$-2.0)}, with most stars
having [Sr/Fe] or [Ba/Fe] $\leq$0.0~\citep{mcw98, bur00, hon04}.  At
higher metallicities, the abundances are within $\sim$0.5 of the solar
abundance.

A cool field giant, U Aquarii, has enhanced [Sr/Fe] and [Y/Fe]
abundances, and low [Ba/Fe] \citep{bln79}.  This star is a faint R CrB
type variable star and shows no CH features, but strong $^{12}$C$_{2}$
bands.  \citet{bln79} concluded that U Aqr is a hydrogen deficient
carbon star with enhanced abundances of the light s-process elements
Sr and Y (by a factor of $\sim$100) and little or no Ba.  It is now a
He-C core of an evolved star of $\sim$1M$\odot$ that ejected its H
rich envelope at the He core flash. \citet{bln79} postulated that a
single neutron exposure occurred at the flash resulting in a brief
neutron irradiation producing only the light s-process elements. A
similar giant, known as Sakurai's Object \citep{asp97} shows similarly
enhanced C and light s-process elements, and is H deficient.  These
types of stars may be responsible for the abundance pattern found in
star 2015448.  However, a significant difference is the C-rich nature of
the giants compared with the carbon depleted nature of star 2015448.

Our results provide a challenging puzzle to determine the source of
the abundance patterns found for this star.  That said, the resolution
of our data is inadequate to address this question.  Higher resolution
spectra with high S/N are needed to analyze as many elements as
possible, in particular the s-process ones, to be able to obtain an
accurate history for the evolution of star 2015448.

\clearpage

\clearpage

\clearpage

\begin{thebibliography}{}
\bibitem[Asplund et al.(1997)]{asp97}
Asplund, M., Gustafsson, B., Lambert, D. L., \& Kameswara Rao, N. 1997, A\&A, 321, L17
\bibitem[Beers et al.(1999)]{bee99} Beers, T. C., Rossi, S., Norris,
J. E., Ryan, S. G., \& Shefler, T. 1999, \aj, 117, 981
\bibitem[Bond, Luck \& Newman(1979)]{bln79}
Bond, H. E., Luck, R. E., \& Newman, M. J. 1979, \apj, 233, 205
\bibitem[Brown \& Wallerstein(1993)]{bw93}
Brown, J. A., \& Wallerstein, G. 1993, \aj, 106, 133
\bibitem[Burris et al.(2000)]{bur00} 
Burris, D. L., Pilachowski, C. A., Armandroff, T. E., Sneden, C., Cowan, J. J., Roe, H. 2000, \apj, 544, 302
\bibitem[Cameron(1982)]{cam82}
Cameron, A. G. W. 1982, \apss, 82 123
\bibitem[Cottrell \& Norris(1978)]{cn78}
Cottrell, P. \& Norris, J. 1978, \apj, 221, 893
\bibitem[Cunha et al.(2002)]{cun02}
Cunha, K., Smith, V. V., Suntzeff, N. B., Norris, J. E., Da Costa, G. S., \& Plez, B. 2002, \aj, 124, 379
\bibitem[Dinescu~et~al.(1999)]{din99}
Dinescu, D I., van Altena, W F., Girard, T M., \& López, C E. 1999a, \aj, 117, 277
\bibitem[Fenner et al.(2004)]{fen04}
Fenner, Y., Campbell, S. W., Karakas, A. I., Lattanzio, J. C., Gibson, B. C. 2004, \mnras, 353, 789
\bibitem[Hilker~et~al.(2004)]{hil04}
Hilker, M., Kayser, A., Richtler, T., \& Willemsen, P.  2004, \aap, 442, L9
\bibitem[Honda et al.(2004)]{hon04}
Honda, S., Aoki, W.,  Kajino, T., Ando, H., Beers, T. C., Izumiura, H.,. Sadakane, K., \& Takada-Hidai, M. 2004, \apj, 607, 474
\bibitem[Lattanzio \& Karakas(2001)]{lk01}
Lattanzio, J. C., \& Karakas, A. I. 2001,  Mem. Soc. Astron. Italiana, 72, 255
\bibitem[Lee et al.(1999)]{lee99}
Lee, Y.-W., Joo, J.-M., Sohn, Y.-J., Rey, S.-C., Lee, H.-C., Walker, A. R. 1999, Nature, 402, 55
\bibitem[Lewis et al.(2002)]{lew02}
Lewis et al., 2002, \mnras, 333, 279
\bibitem[Lloyd Evans(1977)]{le83}
Lloyd Evans, T. 1983, \mnras, 204, 975
\bibitem[Lub(2001)]{lub02}
Lub, J., 2002, in van Leeuwen, F., Hughes, J. D., \& Piotto, G. 2002, eds, ASP Conf. Ser. Vol. 265, A Unique Window into Astrophysics. (Astron. Soc. Pac.: San Francisco), p. 95
\bibitem[McWilliam (1998)]{mcw98}
McWilliam, A. 1998, \aj, 115, 1640
\bibitem[Norris \& Da Costa(1995a)]{nd95a}
Norris, J. E., \& Da Costa, G. S.  1995, ApJ, 447, 680
\bibitem[Norris \& Da Costa(1995b)]{nd95b}
Norris, J. E., \& Da Costa, G. S.  1995, \apj, 441, L81
\bibitem[Norris, Freeman \& Mighell(1996)]{nfm96} 
Norris, J. E., Freeman, K. C., \& Mighell, K. J.  1996, ApJ, 462, 241
\bibitem[Pancino et al.(2000)]{pan00}
Pancino, E., Ferraro, F. R., Bellazzini, M., Piotto, G., \& Zoccali, M.  2000, ApJ, 534, 83
\bibitem[Pancino et al.(2002)]{pan02} 
Pancino, E., Pasquini, L., Hill, V., Ferraro, F. R., Bellazzini, M. 2002, \apj, 568, L101
\bibitem[Persson et al.(1980)]{per80}
Persson, S. E., Frogel, J. A., Cohen, J. G., Aaronson, M., \& Matthews, K. 1980, \apj, 235, 452
\bibitem[Prantzos, Hashimoto \& Nomoto(1990)]{phn90}
Prantzos, N., Hashimoto, M., \& Nomoto, K. 1990, A\&A, 234, 211
\bibitem[Raiteri et al.(1993)]{rai93}
Raiteri, C. M., Gallino, R., Busson, M., Neuberger, D., \& Kappeler, F. 1993, \apj, 419, 207
\bibitem[Rey~et~al.(2004)]{rey04}
Rey, S.-C., Lee, Y.-W., Ree, C. H., Joo, M.-J., \& Sohn, Y.-J. 2004, \aj, 127, 958
\bibitem[Sollima~et~al.(2005)]{sol05}
Sollima, A., Pancino, E., Ferraro, F. R., Bellazzini, M., Straniero, O., Pasquini, L. 2005, \apj, 634, 332
\bibitem[Smith(2005)]{smi05}
Smith, V. V. 2005,  in Barnes, T. G. \& Bash, F. N., eds, ASP Conf. Ser. Vol. 336, Cosmic Abundances as Records of Stellar Evolution and Nucleosynthesis in honor of David L. Lambert. (Astron. Soc. Pac.: San Francisco), 165
\bibitem[Smith et al.(1995)]{scl95}
Smith, V. V., Cunha, K., \& Lambert, D.  1995, AJ, 110, 2827
\bibitem[Smith et al.(2000)]{smi00}
Smith, V. V., Suntzeff, N., Cunha, K., Gallino, R., Busso, M., Lambert, D., \& Straniero, O.  2000, AJ, 119, 1239
\bibitem[Stanford et al.(2006a)]{sta06a}
Stanford, L. M., Da Costa, G. S., Norris, J. E., \& Cannon, R. D. 2006a, \apj, 647, 1075
\bibitem[Stanford et al.(2006b)]{sta06b}
Stanford, L. M., Da Costa, G. S., Norris, J. E., \& Cannon, R. D. 2006b, in preparation
\bibitem[Sunztzeff \& Kraft(1996)]{sk96}
Suntzeff, N. B., \& Kraft, R. P. 1996, \aj, 111, 1913
\bibitem[Truran et al.(2002)]{tru02}
Truran, J. W., Cowan, J. J., Pilachowski, C. A., \& Sneden, C. 2002, PASA, 113, 1293
\bibitem[Ventura et al.(2002)]{vdm02}
Ventura, P., D'Antona, F., \& Mazzitelli, I.  2002, A\&A, 393, 215
\bibitem[Wanajo et al.(2003)]{wti03}
Wanajo, S., Tamamura, M., \& Itoh, N. 2003, \apj, 593,968
\bibitem[Woosley \& Hoffman(1992)]{wh92}
Woosley, S. E., \& Hoffman, R. D. 1992, \apj, 395, 202
\bibitem[Yi et al.(2001)]{yi01}
Yi, S., Demarque, P., Kim, Y.-C., Lee, Y.-W., Ree, C.-H., Lejeune, T., \& Barnes, S. 2001, \apj, 136, 417
\end{thebibliography}
\end{document}